\documentclass[twocolumn,showpacs,showkeys]{revtex4}
\usepackage[dvips]{graphicx}
\usepackage{epsfig}
\begin{document}

\hspace{14cm}\vspace{1cm}

\title{Chaotic Tunneling in a Laser Field }

\author{V.L. Golo$^1$ and Yu.S.Volkov$^1$}
\email{golo@mech.math.msu.su }

\affiliation{$^1$ Department of Mechanics and Mathematics \\
     Moscow University \\
     Moscow 119 899   GSP-2, Russia \\  }

\date{April 3, 2005}

\begin{abstract}
We study  the driven tunneling of a one-dimensional charged
particle confined to a rectangular double-well. The numerical
simulation of the Schr\"odinger equation based on the
Cranck-Nicholson finite-difference scheme, shows that the
modulation of the amplitude of the external field may result in
the parametric resonance. The latter is accompanied by the
breakdown of the quasi-periodic motion characteristic of the usual
driven tunneling, and the emergence of an irregular dynamics. We
describe the above breakdown  with the occupation probability for
the ground state of the unperturbed system, and   make the
visualization of the irregular dynamics with the help of
Shaw-Takens' reconstruction of the state-space. Both approaches
agree as to the values of the resonant frequency for the
parametric excitation. Our results indicate that the shape of the
laser pulse could be essential for generating chaotic tunneling.

\end{abstract}


\maketitle

{\bf 1.}  Driven transitions in a double well are instrumental for
studying the tunneling in various fields of physics and chemistry,
\cite{mont},  \cite{milena}. Considerable attention has been drawn
to the tunneling dynamics in the presence of a driving force with
a time dependent amplitude. In his seminal paper, \cite{holthaus},
M. Holthaus showed that shaping the driving force may be
instrumental in controlling the tunneling in a bistable potential.
In particular, it was shown, \cite{holthaus}, that choosing an
appropriate envelope for laser pulse, one may perform the
population transfer on time scales much shorter than the base
tunneling time. This situation is intimately related to the
problem of quantum chaos, which is generally approached within the
framework of the quasi-classical approximation and Gutzwiller's
theory. In fact, classical chaotic systems are often used as a
clue to the quantum ones. In contrast, it would be very
interesting to look at the quantum chaos the other way round and
consider systems which need studying without approximations that
could have bearing upon classical mechanics, for example,
particles confined to potentials of a size comparable with the de
Broglie wave length. This has also an additional interest owing to
the fact that calculations within the framework of semiclassical
theory should depart from the quantum ones on the time scale of
$\hbar / \Delta E$, $\Delta E$ being the typical spacing between
energy levels.  Therefore, the Schr\"odinger equation describing
the problem needs numerical studying.

\begin{figure}
  \begin{center}
    \includegraphics[width = 7.5cm, height=4.94cm]{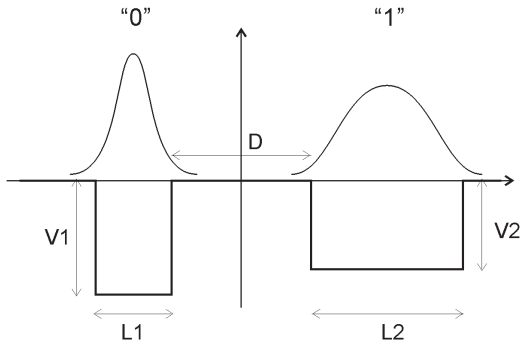}
    \caption{Rectangular one-dimensional double-well potential.
             Lines "0" and "1" indicate the moduli of the wave functions
             $|\psi_0(x,t)|$ and $|\psi_1(x,t)|$ of the ground
             and the first excited states.
             In the chosen configuration $D = 0.876$ is
             the distance between the wells;
             $L_0 = 2.337 , L_1 = 2.045$ are the widths, and
             $U_0 = 13.82 , U_1 = 11.91 $ are the depths
             of the shafts in  dimensionless units.}
             \label{fig1}
  \end{center}
\end{figure}

This letter aims at the theoretical description of the tunneling
transitions for a charged particle  in a one-dimensional
rectangular double-well potential $U(x)$ and an external periodic
field $V(t)$ that mimics electro-magnetic irradiation. The
Hamiltonian of the system in dimensionless units is given by the
equation
$$
    H =  \frac{1}{2} \, \hat{p}^2  + U(x)  + V(t)
$$
in which the potential $U$ reads
\begin{equation}
    U = \left\{ \begin{array}{rl}
                    - U_1, & a \le x \le b \\
                    - U_2, & c \le x \le d \\
                    0,     & \mbox{otherwise}
                \end{array}
        \right.
         \label{potential}
\end{equation}
i.e. the unsymmetrical double well comprising the two shafts of
different depths $U_1, U_2$,  see FIG.\ref{fig1}. The shape of $U$
is chosen in such a way that the values of the wave functions,
$\psi_0(x,t) = <x|0>, \quad \psi_1(x,t) = <x|1>$, of the ground
and the first excited state, are very small outside the right and
the left shafts of the well, respectfully, see FIG.\ref{fig1}.
Consequently, the particle's transitions $|0>
\rightleftharpoons|1>$ can be visualized as the tunneling ones.
The field frequency $\Omega$ verifies the constraint of resonance,
which in dimensionless units reads $\Omega = E_1 - E_0$, where
$E_0, \, E_1$ are the energies of the ground and the first excited
states. The external field $V(t)$ has the form
$$
   V(t) = -  A\, \sin(\Omega t) \, \hat{p}, \quad
   \hat{p} = - i \partial_x
$$
where $\hat{p}$ is the operator of momentum, corresponding to an
electro-magnetic wave described by the vector potential $A_x = A_z
= 0, A_y = A \sin\Omega (t - x / c) $,  the scalar potential
$\phi$ being equal to zero, \cite{scully}. In what follows we
neglect the dependence of $A$ on $x$ since the wave length is
assumed to be much larger than the well's size.

We are looking for regimes in which the regular tunneling breaks
down and becomes chaotic. The main instrument employed for this
end, is the Shaw-Takens (ST) method, \cite{shtk}, for
reconstructing the state-space of a system from a stream of data,
obtained with the numerical integration of the Schr\"odinger
equation, by embedding it in a phase space of enough dimension
${\cal D}$, in our case ${\cal D}=3$. The important thing is that
one may monitor the dynamics of driven tunneling, by using the
occupation probability $N_0$ of the ground state of the
unperturbed system
$$
        N_0 = |<0|\psi>|^2,
$$
where $|0>$ and $|\psi>$ are the ground vector, and the state
vector of the system at the moment of time $t$, respectively. To
obtain the visualization  we shall form a time series that
comprises the values of $N_0$ at moments of time $n \tau, \quad
n=0,1,2, \ldots ,$; $\tau$ being the lag-time, so that $N_{0k} =
N_{0} (k\tau), \, k=1,2,3, \ldots $ . Then the series of vectors $
Y_k = (N_{0 k}, N_{0 k+1}, N_{0 k+2}), k=0,1,2,3, \ldots $, serves
a 3d-visualization of the state-space for the given problem. In
particular,  we can observe the transition from the regular motion
characteristic of the unperturbed resonant motion to the irregular
one generated by a parametric excitation given by
Eq.(\ref{parametric2}) below, see FIG.\ref{thorus}.

{\bf 2.} The  parametric excitation of the system is generated by
the modulation of the amplitude $A$ of the vector potential
\begin{equation}
   A = A_0 [1 - \epsilon \sin(\omega t)]
   \label{parametric2}
\end{equation}
The approximate expression for the frequency of the parametric
resonance caused by (\ref{parametric2}) can be obtained by
employing the rotating wave approximation, \cite{scully} after
some analytical calculations. It reads
\begin{equation}
     \omega_{prm} =  A \kappa, \quad \kappa = |<0|\hat{p}|1>|
   \label{rotwave}
\end{equation}
In fact, the actual resonant frequency $\omega_{prm}$ differs from
that given above and its calculation requires some numerical work.

{\bf 3.}  The visualization with the help of the ST-method gives a
torus in 3d-space, fairly well drawn,
    changes in the lag-time resulting only in continuous
    deformations of the picture, and no sudden modifications observed,
    see FIG.\ref{thorus}.
    In contrast, in case  the parametric excitation at a frequency
    close to the value given by (\ref{rotwave}), being present,
    the choice of the lag-time is very important for obtaining a
    meaningful visualization. The break down of the quasi-periodic
    motion by the resonant parametric excitation  is shown in FIG.\ref{thorus}.
It is to be noted that the visualization depends on a number of
premises, and first of all the value of the lag-time $\tau$ and
the dimension $d$ of the visualization window. The wise choice of
$\tau$ is dictated by characteristic times of the motion under the
investigation, so that for certain values of $\tau$ the
visualization picture is coherent enough whereas for others it is
not meaningful. Therefore, using the ST-method one has to compare
pictures obtained for different values of the lag-time.

{\bf 4.} We may look at our problem the other way round. Consider
sets of the system's states ${\cal F}_{k k+1}, \, k=1,2, \ldots ,
{\cal N}$, where ${\cal N}$ is large enough,  given by constraints
\begin{equation}
       \Delta_k \le N_0 \le \Delta_{k+1},
       \label{folio}
\end{equation}
where $\Delta_k$ are intervals dividing segment $0 \le N_0 \le 1$
in ${\cal N}$ equal parts. Take a period of time $T$, sufficiently
large, and consider times $\tau_k$ spent by the system in the sets
of states ${\cal F}_{k k+1}$, that is where (\ref{folio}) is
verified. We shall define the visiting frequencies as
$$
      \nu_{k k+1} = \frac{\tau_k}{T}
$$
One may cast the $\nu_{k k+1}$  in the form of the probability
density, $\xi(N_0)$, for the distribution  of $N_0$, by assuming
that all the intervals $\Delta_{k+1} - \Delta_k$ be of equal size
$\Delta$, and defining
$$
    \xi(N_0) = \frac{\nu_{k k+1}}{\Delta}
$$
On considering the limit of $\xi$ as ${\cal N} \to \infty$, we
shall get the probability density $\xi(N_0)$, see FIG.\ref{fig3}.

To put it in a quantitative analytical form, we may consider the
characteristic function  $ \chi_{ab}(x) = 1, \mbox{if} \quad a \le
x \le b$ and $0$ otherwise, and introduce the quantity
\begin{equation}
    \nu_{\phi}(a,b) = \lim_{T \rightarrow \infty}
                      \left\{ \frac{1}{T} \int_{-T/2}^{+T/2} \,
                              \chi_{ab}(N_0) \, dt
                      \right\},
        \label{vfreq}
\end{equation}
so that $\nu_{ab}$ can be considered as the frequency of visiting
a region of states determined by constraint $a \le N_{\phi} \le
b$.

The important point is that the limit indicated in the above
equation does exist in the context of the problem under
investigation. To see the fact we shall employ the normalization
of wave functions in a finite box, that is,  our system being
confined to a rectangular potential of infinite depth of a size
much larger than the double well's size. Then the system has the
discrete spectrum of eigenvalues, and the characteristic function
$\chi_{ab}(N_0)$ can be expanded in Fourier series,  the time
averaging in the above equation results in cancelling out terms
oscillating in time, so that  the limit exists. The use of the
finite box normalization is also important for managing the
numerical simulation. It is alleged to be known, the finite
integration mesh, or basis, results in
reflection of the wave packet that distorts the time evolution of
the packet. By imposing  the finite box normalization we use the
physical framework that appears to be compatible with the
tunneling dynamics in the double well of a size much smaller than
the box and enables us to overcome the numerical artifacts.


It is worth noting that the construction of visiting frequencies
is analogous to that of the dwell time which is aimed at studying
the localization of wave packets and defined by the equation,
\cite{buttiker}, \cite{nuss},
$$
    \tau_D(a,b) = \int^{+\infty}_{-\infty} \,dt \, \int^b_a \,
                  |\psi(x,t)|^2 \, dx
                  \label{dwell}
$$

\begin{figure}
  \begin{center}
    \includegraphics[width = 7.5cm, height=6.47cm]{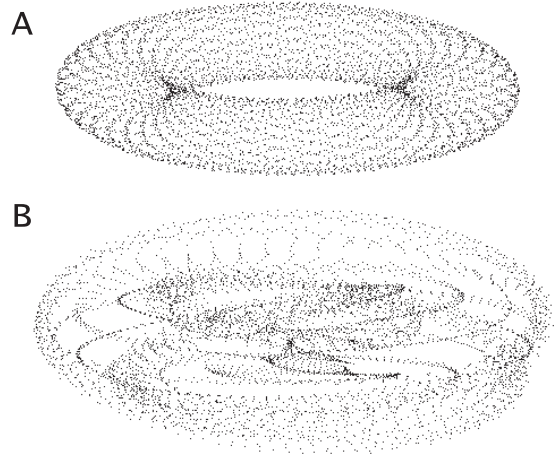}
    \caption{Visualization of the state space.
             {\bf A} no parametric excitation.
             {\bf B} parametric excitation:
                $A_0 = 0.3; \epsilon = 0.1; \omega = 0.01755 $.
                The lag-time $330$, in both cases.}
             \label{thorus}
  \end{center}
\end{figure}

The probability density $\xi$ indicates that at the resonant
frequency $\omega_{prm}$ the character of the system's motion
undergoes a drastic change, see FIG.\ref{fig3}. At this point it
should be noted that the frequency of the parametric resonance
$\omega_{prm}$ which can be found  by comparing the sizes of the
peaks in FIG.\ref{fig3} for different values of $\omega$ in equation
(\ref{parametric2}),   differs by approximately $5\%$ from the
value given by the rotation wave approximation (\ref{rotwave}).
The resonant value corresponds to the most pronounced breakdown of
the twin peaks, which correspond to the ground and the first
excited state of the free system.

\begin{figure}
  \begin{center}
    \includegraphics[width = 7.5cm, height=7.08cm]{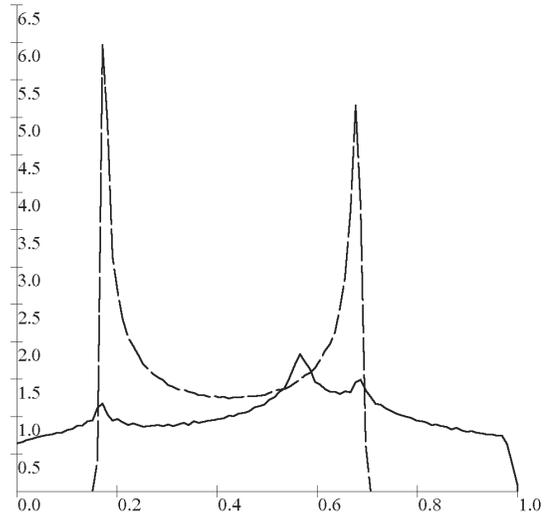}
    \caption{Probability density $\xi$ for the distribution of $N_0$
             with $A_0=0.3$;  6 six bound states (levels confined to the well).
             Dashed line - no parametric excitation.
             Solid line - parametric excitation with  $\epsilon = 0.1$.,
             parametric frequency $\omega = 0.01755 $ close
             to the resonant one, $\omega_{prm}$}
             \label{fig3}
  \end{center}
\end{figure}

{\bf 5.} Concluding we should like to note that, as  follows from
Eq.(\ref{parametric2}) for the amplitude of the driving force,
the pulse $V$ has a triplet structure determined by the main
contribution at frequency $\Omega = E_1 - E_0$ and two satellites
at $\Omega \pm \omega_{prm}$, of less amplitude, owing to the
equation for $A(t)$ given by (\ref{parametric2})
\begin{eqnarray*}
    A(t) &=&  A_0 \kappa \,  \, \sin \Omega t
         - \frac{1}{2} A_0\kappa \,  \epsilon \,
           \cos [(\Omega - \omega_{prm})t] \\
          &+& \frac{1}{2} A_0\kappa \,  \epsilon \, \cos [(\Omega + \omega_{prm})t]
\end{eqnarray*}

The triplet structure could generally have an important bearing on
the tunneling in the double-well potential.  In fact, if a
monochromatic pulse at resonant frequency $\Omega = E_1 - E_0$ is
employed, there is no parametric excitation and the dynamics of
the occupation probability has the usual form. In contrast, a poor
quality non-monochromatic pulse may contain the triplet $\Omega
\pm \omega_{prm}$ so that the deformation of the tunneling
dynamics and the emergence of the chaotic motion become possible.
It is worth noting that the amplitude of the driving force, $A_0$,
can be small so that the frequency shift determined by
$\omega_{prm}$ be tiny.

\noindent {\bf Acknowledgment} \\
This work was supported by the grants NS - 1988.2003.1, and RFFI
01-01-00583, 03-02-16173, 04-04-49645.

\end{document}